\documentclass[conference]{IEEEtran}
\IEEEoverridecommandlockouts
\usepackage{cite}
\usepackage{amsmath,amssymb,amsfonts}
\usepackage{algorithmic}
\usepackage{graphicx}
\usepackage{subcaption} 
\usepackage{textcomp}
\usepackage{xcolor}
\usepackage{booktabs}
\usepackage{multirow}
\usepackage{algorithm}
\usepackage{tcolorbox}
\usepackage{soul}

\newcommand{\INPUT}{\item[\textbf{Input:}]}

\newcommand{\OUTPUT}{\item[\textbf{Output:}]}

\def\BibTeX{{\rm B\kern-.05em{\sc i\kern-.025em b}\kern-.08em
    T\kern-.1667em\lower.7ex\hbox{E}\kern-.125emX}}
\begin{document}

\title{ALPHA: LLM-Enabled Active Learning for Human-Free Network Anomaly Detection \\
\author{
		\IEEEauthorblockN{Xuanhao Luo,
        Shivesh Madan Nath Jha,
        Akruti Sinha,
        Zhizhen Li,
        Yuchen Liu
 }
		\IEEEauthorblockA{
		North Carolina State University, USA
		}
}

}

\maketitle

\begin{abstract}
Network log data analysis plays a critical role in detecting security threats and operational anomalies. Traditional log analysis methods for anomaly detection and root cause analysis rely heavily on expert knowledge or fully supervised learning models, both of which require extensive labeled data and significant human effort. To address these challenges, we propose ALPHA, the first \underline{\textbf{A}}ctive \underline{\textbf{L}}earning \underline{\textbf{P}}ipeline for \underline{\textbf{H}}uman-free log \underline{\textbf{A}}nalysis. ALPHA integrates semantic embedding, clustering-based representative sampling, and large language model (LLM)-assisted few-shot annotation to automate the anomaly detection process. The LLM annotated labels are propagated across clusters, enabling large-scale training of an anomaly detector with minimal supervision. To enhance the annotation accuracy, we propose a two-step few-shot refinement strategy that adaptively selects informative prompts based on the LLM's observed error patterns.
Extensive experiments\textsuperscript{1} on real-world log datasets demonstrate that ALPHA achieves detection accuracy comparable to fully supervised methods while mitigating human efforts in the loop.  ALPHA also supports interpretable analysis through LLM-driven root cause explanations in the post-detection stage. These capabilities make ALPHA a scalable and cost-efficient solution for truly automated log-based anomaly detection.

\footnotetext[1]{{The source code of our proposed ALPHA framework is available at \textit{https://github.com/Xuanhao-Luo/ALPHA}.}}
\end{abstract}


\section{Introduction}
Log data analysis is a fundamental process for detecting security threats, diagnosing system failures, and ensuring operational efficiency in modern computing environments \cite{he2021survey}. Network system logs typically capture detailed records of system behavior, network performance, and security events, providing valuable insights for monitoring and analysis. Anomalous log entries can signal potential cyberattacks, software malfunctions, or network failures, making timely and accurate anomaly detection essential for maintaining system reliability. However, the increasing scale and complexity of modern infrastructures, such as large-scale cloud systems, distributed computing environments, and mobile networks, pose significant challenges to traditional human-based log analysis~\cite{landauer2023deep}, as manual inspection cannot keep pace with the sheer volume, diversity, and velocity of logs generated by these systems.
This, in turn, limits the effectiveness of log-based anomaly detection, as existing methods typically rely on supervised learning approaches that require large amounts of manually labeled data—an approach that is both costly and impractical, particularly when anomalies are rare, system configurations frequently change, or new failure modes arise. Additionally, log semantics are often highly domain-specific, making it difficult to generalize annotation efforts across diverse systems.

Recent advances in large language models (LLMs)
have demonstrated strong capabilities in understanding natural language, making them promising tools for processing and analyzing the network log data, which often contains rich textual descriptions of system events. However, network logs also contain highly structured and domain-specific elements, such as IP addresses, file paths, and error codes, that general-purpose LLMs are not inherently equipped to interpret without task-specific adaptation. Furthermore, LLM inference is computationally intensive, rendering the direct application of LLMs for end-to-end log anomaly detection impractical in resource-constrained environments like edge or mobile devices. To address these limitations, 
we propose integrating LLMs with cost-efficient classifiers to achieve both practicality and high performance. This approach leverages the LLMs’ ability to interpret the structured and semantic features of log data, enabling them to function as accurate \textbf{data annotators} prior to classification and as \textbf{insightful analyzers} for post-detection interpretation. Specifically, we integrate LLMs into an active learning workflow~\cite{settles2009active} to enhance model performance with minimal labeled data by prioritizing the most informative log samples for annotation. Furthermore, by leveraging their inherent reasoning capabilities, LLMs can perform zero-shot root cause analysis of network anomalies, significantly reducing reliance on domain experts for manual intervention.

To this end, we propose \textbf{ALPHA}, an \underline{\textbf{A}}ctive \underline{\textbf{L}}earning \underline{\textbf{P}}ipeline for \underline{\textbf{H}}uman-free log \underline{\textbf{A}}nalysis, enabling fully automated network anomaly detection and interpretation, with no human in the loop--eliminating human intervention across the labeling, training, and explanation stages of log data processing. Labels are automatically generated by LLMs and propagated through clustering based on semantic embeddings, while anomalies are detected by classical machine learning (ML) models and interpreted using prompt-driven LLMs.
Particularly, unlike traditional uncertainty-based active learning strategies that rely on uncertain samples for annotation, ALPHA adopts a novel clustering-based sampling mechanism to select representative log samples, enabling efficient pseudo-label propagation and minimizing human labeling efforts, thereby generating high-quality labeled datasets at scale.
The main contributions of this work are summarized as follows:

\begin{itemize}
    \item We propose ALPHA, a fully automated, human-free framework for log anomaly detection that integrates LLMs into an active learning pipeline for labeling, model training, and network anomaly interpretation.
    
    \item We develop a clustering-based active sampling strategy that selects representative log messages for LLM annotation and propagates these annotations to semantically similar logs, significantly improving the \textbf{efficiency} of labeling automation in practice.
    
    \item We design a two-step few-shot prompting method to enhance LLM-based annotation \textbf{accuracy} by adaptively selecting effective few-shot examples based on inherent error patterns and semantic diversity of log data.
    
    \item Extensive experimental results demonstrate that ALPHA achieves performance comparable to fully supervised baselines, without requiring human-annotated training data. By integrating LLMs into the analysis process, our approach also enhances the interpretability of detected anomalies, enabling truly automated investigation and informed decision-making.
\end{itemize}

\section{Related Works}

\textbf{Conventional Clustering Approaches.} There are traditional ML approaches for log-based anomaly detection. For instance, log clustering structures and groups raw logs into templates or event types for downstream security analysis \cite{landauer2020system}, without requiring labeled data or semantic models.
In \cite{vaarandi2003data}, the authors propose a clustering-based algorithm that mines frequent line patterns and detect outliers from log files, which assists in building system profiles and identifying potential anomalies without pre-labeled data. Similarly, \cite{he2017drain} proposes an efficient online log parsing method that leverages a fixed-depth parse tree to structure raw log messages in a streaming fashion, improving the accuracy and scalability of downstream log analysis tasks like anomaly detection.
Furthermore, \cite{landauer2018dynamic} presents a dynamic, unsupervised cluster evolution method for detecting anomalies in system logs, which incrementally clusters log lines over sliding time windows, monitors cluster changes, and applies time-series forecasting to detect anomalies based on cluster evolution metrics. However, while these clustering methods can structure logs without labeled data, they often rely on shallow features and still require manual inspection for accurate anomaly detection, limiting their scalability and effectiveness in reducing human workload.

\textbf{From Deep Learning to Self-Supervised Framework.} Recent years, many deep learning-based methods have been proposed for log anomaly detection \cite{landauer2023deep}, leveraging architectures such as Long Short-Term Memory networks (LSTMs), Convolutional Neural Networks (CNNs), and Transformers to learn complex patterns from log sequences. 
For instance, \cite{zhang2019robust} proposes LogRobust, which leverages semantic vector representations and an attention-based Bi-LSTM model to perform robust anomaly detection on evolving and noisy log data.
\cite{le2021log} proposes a Transformer-based framework that leverages BERT to directly learn semantic representations from raw log messages, enabling effective log anomaly detection without relying on log parsing.
\cite{guan2024logllm} introduces a dual-LLM framework combining BERT and LLaMA for log-based anomaly detection, leveraging regex-based preprocessing and a three-stage training strategy to align representations and improve detection performance across diverse datasets.
Similarly, \cite{guo2021logbert} proposes LogBERT, a BERT-based self-supervised framework for log anomaly detection, leveraging masked log key prediction and hypersphere minimization to model normal sequences and identify anomalies without requiring labeled data.
While existing self-supervised methods such as DeepLog and LogBERT reduce reliance on labeled anomaly data, they still assume access to a clean set of normal logs, implicitly requiring prior extensive human annotation. 

\textbf{Towards Active Learning Strategy.} A few works have explored active learning strategies to reduce such annotation costs in log anomaly detection tasks.
\cite{duan2023afalog} introduces an active learning-based framework AFALog to enhance unsupervised anomaly detection by addressing Not-Cover problem and the Suspicious-Noise problem, relying on normal logs while incorporating selectively annotated samples for improved accuracy.
In \cite{xiao2024logcae}, authors leverage active learning and contrastive learning to enhance log-based anomaly detection by selectively labeling uncertain samples and refining feature representations, reducing reliance on extensive annotations.
Both methods rely on entropy-based active learning strategies to select samples for annotation. However, they still require labeling approximately 6\% of the data to achieve a decent result. This remains a significant human effort, especially in scenarios where ML models demand large-scale training data {(tens of thousands of logs)} or log patterns frequently evolve, requiring frequent retraining.
To address this, we innovate a human-free pipeline that combines semantic embeddings, log clustering, and LLM-based few-shot annotation, enhanced by active learning workflow to eliminate manual labeling while maintaining a high detection quality.

\section{Motivation and Observations}
\subsection{Motivation for Active Learning}
\begin{figure}[t]
	\centerline{\includegraphics[scale=0.7]{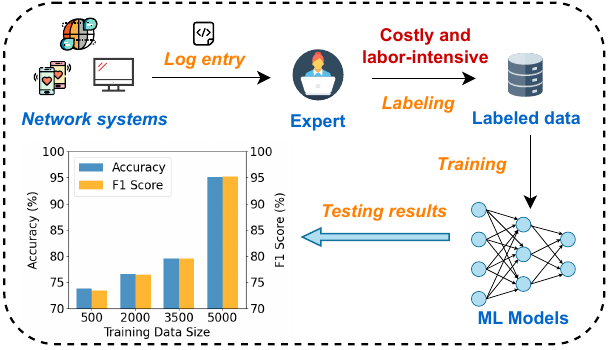}}
    \caption{Traditional training process for log anomaly detection.} 
    \label{motivation}
    \vspace{-5mm}
\end{figure}

Traditional supervised ML approaches to log anomaly detection require large amounts of labeled data. As illustrated in Fig.~\ref{motivation}, system-generated network logs are manually labeled by domain experts before being used to train ML-based anomaly detection models. 
While the scaling law~\cite{gao2020consistency} suggests that increasing the volume of labeled data generally improves model performance, as demonstrated in our preliminary study using a Support Vector Machine (SVM) classifier (bottom left of Fig.~\ref{motivation}), where anomaly detection accuracy improves from 73.80\% to 95.12\% as the training size grows from 500 to 5,000 log samples, the manual labeling process remains both costly and labor-intensive in the loop — requiring nearly a 900\% increase in annotation effort for about a 20\% gain in performance, thus posing a significant bottleneck to scaling data-driven solutions.

\textbf{Observation 1.} \textit{While log anomaly detector performance improves with larger labeled datasets, the cost and effort of manually labeling log data can be prohibitively high.}

This observation underscores the necessity of data-efficient active learning strategies, wherein the model selectively queries the most informative samples for annotation instead of exhaustively labeling the entire dataset. By progressively expanding the labeled dataset in a targeted manner, we aim to minimize annotation costs while maximizing model accuracy.

\subsection{Motivation for Clustering and Label Propagation}

\begin{figure}[ht]
    \centering
    \begin{subfigure}{0.47\linewidth} 
        \centering
        \includegraphics[width=\linewidth]{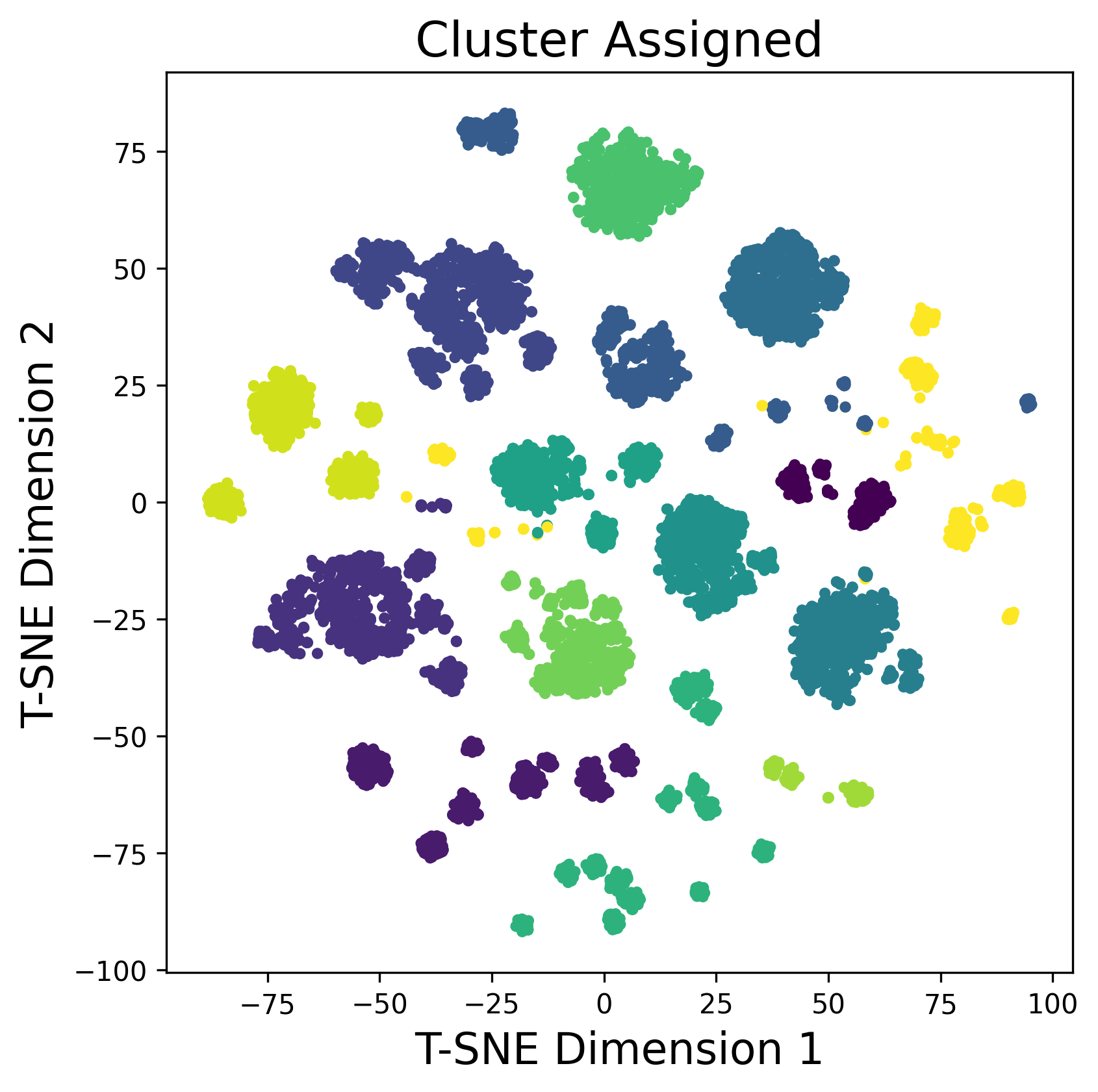}
        \caption{t-SNE visualization of K-Means cluster assignments.} 
	\label{loss}
    \end{subfigure}
    \hfill
    \begin{subfigure}{0.47\linewidth}
        \centering
        \includegraphics[width=\linewidth]{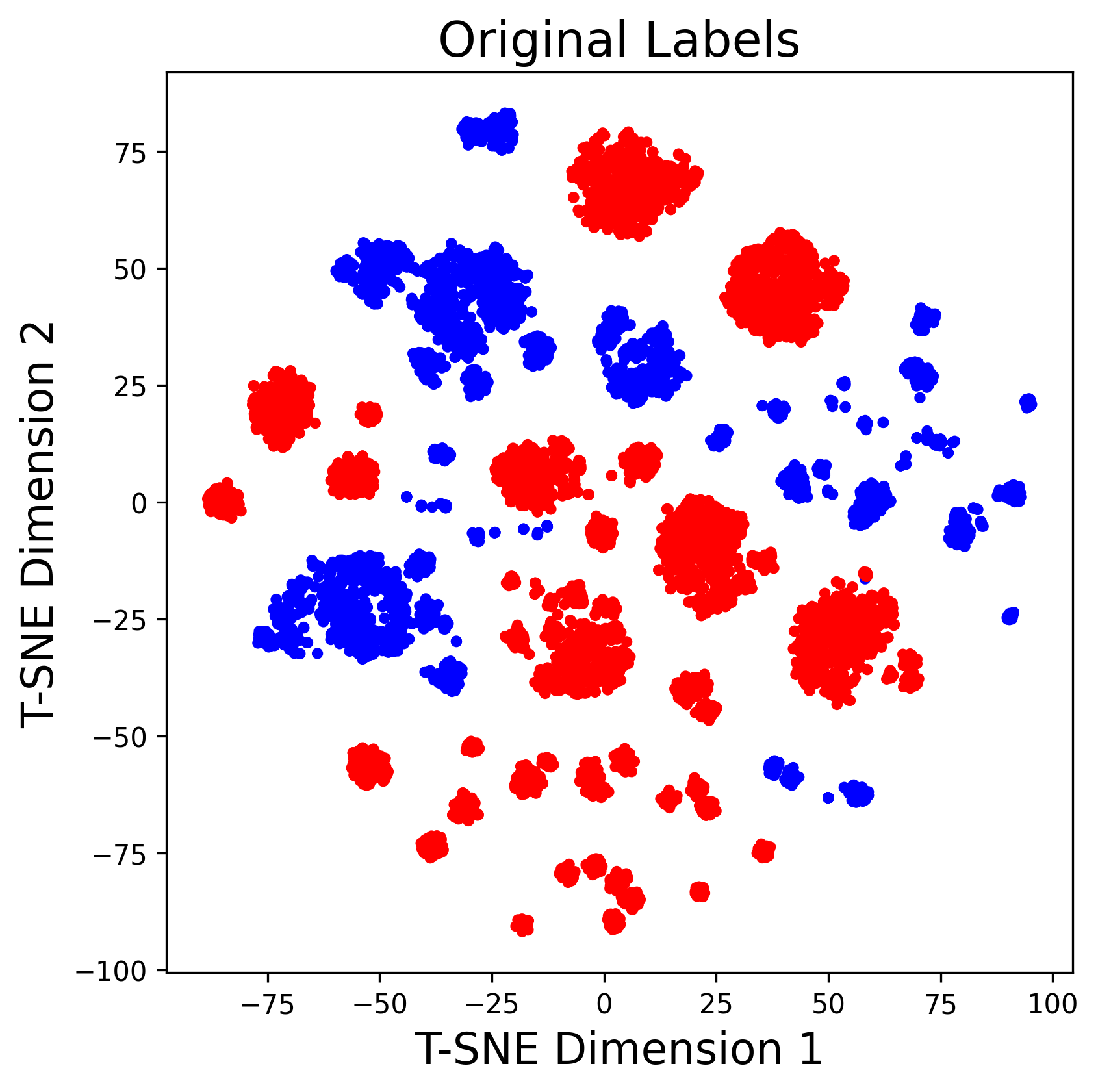}
        \caption{t-SNE visualization of ground-truth labels.} 
	\label{Entropy}
    \end{subfigure}
    \caption{Comparison of the latent space representations: (a) shows cluster assignments obtained via K-Means, and (b) displays the abnormal labels ({red}) vs. normal labels ({blue}).} 
	\label{tsne}
    \vspace{-2mm}
\end{figure}

Log data, as time-series records, inherently exhibit temporal and structural patterns, motivating our adoption of a clustering-based label propagation approach to further reduce annotation costs. To validate this strategy, we conducted a preliminary experiment comparing the t-SNE visualizations of clustered log embeddings (a new representation detailed in Sec. IV.A) against the ground-truth labels (normal or abnormal events). This comparison assesses whether the learned embeddings reveal meaningful structures and if clustering effectively separates normal from anomalous instances. If the embedding space naturally groups semantically similar log entries, instances within the same cluster are likely to share the same label, thus enabling efficient propagation of annotations.
\figurename~\ref{tsne} presents the t-SNE projections of the log embeddings. \figurename~\ref{tsne}(a) illustrates the cluster assignments obtained via K-Means clustering, while \figurename~\ref{tsne}(b) shows the ground-truth labels. The visual correspondence between clusters and true labels leads to the following observation:

\textbf{Observation 2.} \textit{The embeddings exhibit a well-structured distribution where distinct clusters align closely with actual anomalies and normal log events.}

This suggests that the embedding space preserves meaningful semantic relationships, allowing unsupervised clustering to serve as an effective prior for label propagation. By labeling only a few representative samples per cluster and propagating these labels within the cluster, one can substantially reduce manual annotation costs while maintaining detection accuracy.

\begin{figure*}[t]
	\centerline{\includegraphics[scale=0.76]{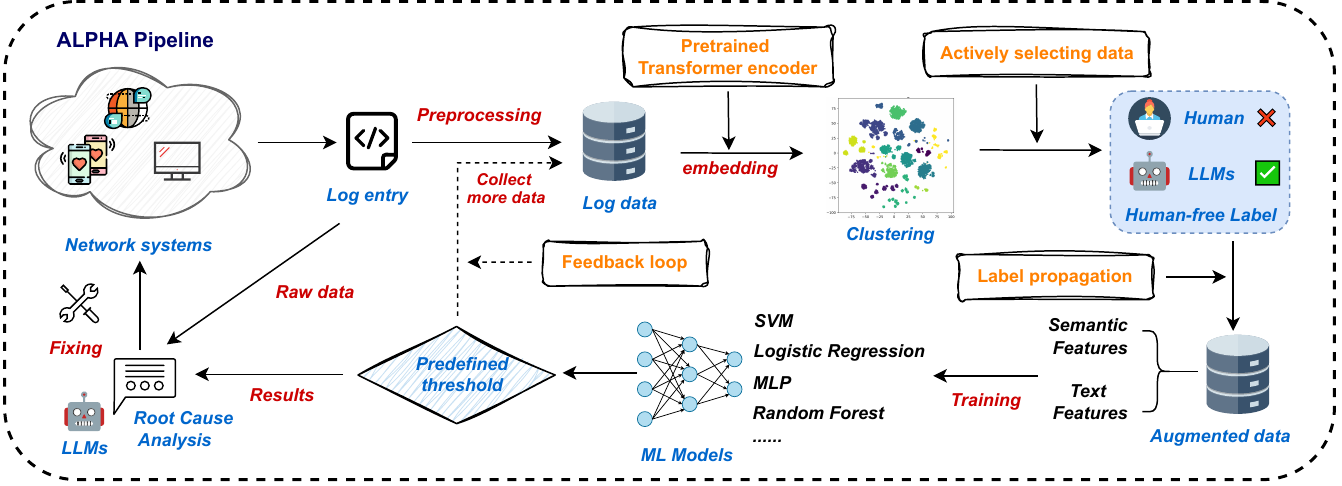}}
    \caption{Overview of ALPHA for closed-loop human-free log analysis.} 
    \label{framework}
    \vspace{-6mm}
\end{figure*}

\subsection{Motivation for LLM-based Annotation}

Motivated by the potential of LLMs in understanding log data, we conducted a preliminary study to investigate whether LLMs can directly serve as a data labeler when provided with raw log inputs. As shown in Table~\ref{llm}, we explored several prompt engineering strategies to enhance annotation accuracy. Specifically, we observed that Few-Shot Learning (FS) achieved an accuracy of 71\% with an F1-score of 76\%. Incorporating Chain-of-Thought (CoT) reasoning alongside FS further improved performance, raising accuracy to 74\% and the F1-score to 78\%. Finally, the combination of FS, CoT, and Self-Reflection (SR) yielded a marginal improvement, reaching the highest accuracy of 75\% and an F1-score of 79\%.
Although LLM-based annotation is not yet perfectly accurate, these results suggest the following observation:

\textbf{Observation 3.} \textit{LLMs exhibit strong potential as automated annotators for log data, especially when equipped with prompt engineering strategies. As FS learning emerges as an effective prompting approach, it opens opportunities to further improve annotation accuracy by incorporating carefully selected few-shot examples and majority voting mechanisms.}

This finding indicates that, although LLMs alone may not fully replace human experts in all cases, the integration of prompt engineering, strategic sample selection, and voting-based aggregation (detailed in Sec. IV-C) could substantially enhances their effectiveness, establishing LLMs as a viable and scalable approach for automated log annotation.

\begin{table}[t]
\centering
\caption{GPT-4o \cite{hurst2024gpt} Labeling Performance.}
\begin{tabular}{c|cc}
\toprule
\textbf{Technique} & \textbf{Accuracy (\%)} & \textbf{F1-score (\%)} \\
\midrule
FS & 71 & 76 \\
FS+CoT (FS first) & 73  & 77 \\
CoT+FS (CoT first) & 74 & 78 \\
FS+CoT+SR & 75 & 79 \\
\bottomrule
\end{tabular}
\label{llm}
\vspace{-5mm}
\end{table}

\section{LLM Assisted Active Learning for Log Data Analysis}

In this section, we introduce the architecture of our proposed ALPHA framework, as illustrated in Fig. \ref{framework}. ALPHA begins with raw log entries generated from network systems, which are preprocessed and transformed into dense embeddings using a pretrained Transformer encoder. These embeddings are then clustered to uncover semantically meaningful structures. Instead of relying on traditional uncertainty-based sampling, ALPHA actively selects representative samples from each cluster for LLM-based few-shot annotation. Cluster labels are determined through a majority voting mechanism and propagated to the remaining log data, forming an augmented labeled dataset. This dataset is then used to train lightweight ML models for anomaly detection. If detection performance falls below a predefined threshold, ALPHA initiates a \textbf{feedback loop} that expands the labeled set at the preprocessing stage and retrains the ML detector. Additionally, ALPHA integrates LLM-based log interpretation to support root cause analysis post-detection. In the following sections, we detail each core technical component of the ALPHA framework.

\subsection{Log Data Embedding}
\label{embedding}
\begin{figure}[t]
\centerline{\includegraphics[scale=0.56]{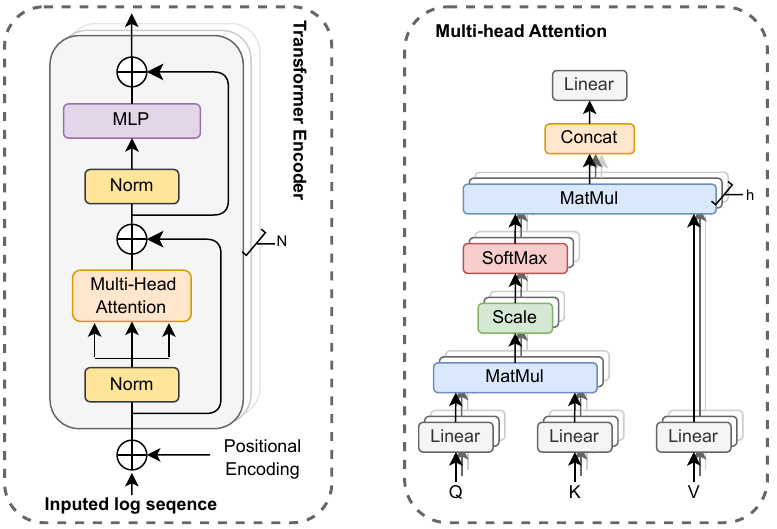}}
    \caption{Transformer encoder for log data embedding.} 
    \label{encoder}
    \vspace{-6mm}
\end{figure}

Transformers, initially developed for natural language processing \cite{vaswani2017attention}, have become a versatile backbone for modeling sequential data across domains, including time-series \cite{xing2025netsight}, network packets \cite{luo2025rank}, and wireless signals \cite{li2025bfmloc}. 
The Transformer encoder architecture, as shown in Fig.~\ref{encoder}, leverages self-attention to capture long-range dependencies across an entire sequence and extract high-quality embeddings.
Unlike decoders that generate sequences autoregressively using causal masking, the encoders process the entire input sequence at once, allowing each token to attend to all others bidirectionally. The final output is a fixed-dimensional vector that encodes the semantic meaning of the log message, making it well-suited for downstream tasks such as semantic clustering and anomaly detection.

Through training such Transformer encoders, we obtain pretrained embedders capable of capturing deep semantic representations of log data. 
By leveraging such pretrained embedders, a set of log messages \( \{L_1, L_2, ..., L_n\} \) can be transformed into embedding vectors as:
\begin{equation}
    E_i = {TransEncoder}_\theta(L_i), \quad \forall i \in \{1, 2, ..., n\},
\end{equation}
where \( E_i \) denotes the embedding for the \( i \)-th log message, and \( {TransEncoder}_{\theta} \) represents the pretrained Transformer encoder parameterized by \( \theta \).

After obtaining the semantic embeddings of log messages via the Transformer encoder, these embeddings are used to drive the active learning process, aiming to efficiently select a small set of informative samples for annotation and propagate high-quality labels across the remaining log data.
Broadly, active learning techniques can be categorized into three classes including query synthesis that creates artificial samples \cite{angluin1988queries}, stream-based sampling that evaluates data as it arrives \cite{cohn1994improving}, and pool-based sampling that selects instances from a static dataset \cite{lewis1995sequential}. Among these, pool-based sampling is particularly suitable for log datasets, as it enables batch selection of representative samples based on global structural properties in the embedding space. 

However, these conventional active learning strategies typically rely on uncertainty-based criteria to prioritize ambiguous samples for annotation. In the context of log data, where many entries exhibit structural or semantic similarities (e.g., recurring error patterns or routine status messages), such uncertainty-driven approaches often result in redundant annotations and inefficient use of labeling resources.
To address this limitation, we propose a clustering-based sampling strategy that groups similar logs together, selects representative samples from each cluster for labeling, and propagates their labels to other samples within the same cluster, as discussed next (in Algorithm 1).

\subsection{Clustering-Based Active Sampling and Label Propagation}

\subsubsection{Log Message Clustering}
Firstly, we employ the k-means algorithm to cluster the LLM-generated embeddings of log messages. Given a set of log embeddings \( \mathcal{E} = \{E_1, E_2, ..., E_n\} \), the objective of k-means is to partition them into \( k \) clusters by the minimizing within-cluster sum of squares (WCSS):
\begin{equation}
    \arg\min_{C} \sum_{i=1}^{k} \sum_{E_j \in C_i} \|E_j - \mu_i\|^2,
\end{equation}
where \( C_i \) represents the set of embeddings assigned to the \( i \)-th cluster, and \( \mu_i \) denotes the centroid of that cluster.

Selecting an appropriate number of clusters \( k \) is crucial for ensuring well-separated and meaningful clusters. In this regard, we employ two widely used schemes to determine the optimal \( k \): the elbow method and silhouette analysis.
The elbow method evaluates WCSS for different values of \( k \). The WCSS generally decreases as \( k \) increases, but the rate of decrease slows down after a certain point, forming an ``elbow'' in the curve. The optimal \( k \) is chosen at the elbow point, where adding more clusters yields diminishing returns in variance reduction.
To further validate the choice of \( k \), we compute the silhouette coefficient, which evaluates the separation and compactness of clusters. The silhouette score for a data point \( E_j \) is given by: 
\begin{equation}
    s(E_j) = \frac{b(E_j) - a(E_j)}{\max(a(E_j), b(E_j))},
\end{equation}
where \( a(E_j) \) is the average intra-cluster distance (i.e., the average distance to other points in the same cluster) and \( b(E_j) \) is the average inter-cluster distance (i.e., the average distance to points in the nearest neighboring cluster). The silhouette coefficient ranges from \(-1\) to \(1\), where higher values indicate better clustering. The optimal \( k \) is selected by maximizing the average silhouette score across all log data samples.
By jointly analyzing the elbow method and silhouette scores, an appropriate cluster number can be determined that balances compactness and separation. It is worth noting that both methods are \textit{unsupervised} and do not require any ground-truth labels, making it easy to determine an appropriate number of clusters (i.e. \textit{k}) when applied to new scenarios or when new data stream arrives.

After determining the optimal cluster number, the k-means clustering process consists of three main stages. First, the centroids \( \mu_1, \mu_2, \dots, \mu_k \) are randomly initialized from the set of log embeddings \( \mathcal{E} \). Each cluster assignment is initialized as empty (Lines 1--3). Then, each embedding \( E_j \) is assigned to the nearest centroid \( \mu_i \) by minimizing the squared Euclidean distance (Line 5). After assignment, each cluster's centroid is updated as the mean of all embeddings within the cluster (Line 6). If the centroids remain unchanged, the algorithm terminates early (Lines 7--9).

\begin{algorithm}[t]
    \caption{Clustering-Based Active Sampling and Label Propagation Algorithm}
    \label{cluster algorithm}
    \begin{algorithmic}[1]
        \INPUT Log embeddings \( \mathcal{E} = \{E_1, E_2, \dots, E_n\} \), number of clusters \( k \), number of representative samples per cluster \( m \), maximum iterations \( T \), confidence threshold \( \tau \)
        \OUTPUT Cluster assignments \( C = \{C_1, C_2, \dots, C_k\} \), Augmented labeled dataset \( \mathcal{D}_{\text{aug}} \)

        \STATE Initialize cluster centroids \( \mu_1, \mu_2, \dots, \mu_k \) randomly from \( \mathcal{E} \);
        \STATE Initialize cluster assignments \( C_i \gets \emptyset \), \( \forall i \in \{1, 2, \dots, k\} \);
        \STATE Set iteration counter \( t \gets 0 \);
        
        \WHILE{\( t < T \)}
            \STATE Assign each embedding \( E_j \) to the nearest centroid:
            \[
            C_i \gets \arg\min_{i} \|E_j - \mu_i\|^2, \quad \forall E_j \in \mathcal{E}
            \]
            
            \STATE Update centroids for each cluster:
            \[
            \mu_i \gets \frac{1}{|C_i|} \sum_{E_j \in C_i} E_j, \quad \forall i \in \{1, 2, \dots, k\}
            \]

            \IF{Centroids remain unchanged}
                \STATE \textbf{Break}; 
            \ENDIF

            \STATE Increment iteration counter \( t \gets t + 1 \);
        \ENDWHILE

        \FOR{each cluster \( C_i \)}
            \STATE Select the \( m \) most representative samples \( S_i \subset C_i \) closest to centroid \( \mu_i \);
            \STATE Obtain labels \( Y_{S_i} \) for selected samples using LLM-based annotation;
            \STATE Compute final cluster label \( Y_{C_i} \) via majority voting over \( Y_{S_i} \);
            \STATE \textbf{Propagate} \( Y_{C_i} \) to all samples in \( C_i \);
            \STATE Add labeled log data pair \( (E_j, Y_{C_i}) \) to \( \mathcal{D}_{\text{aug}} \), \( \forall E_j \in C_i \);
        \ENDFOR

        \STATE \textbf{return} \( C, \mathcal{D}_{\text{aug}} \);
    \end{algorithmic}
\end{algorithm}
\noindent

\subsubsection{Representative Sample Selection and Label Propagation}

Once clusters are formed, a representative sample selection strategy is proposed to improve label robustness and minimize annotation efforts. Instead of selecting only the closest sample to the centroid sample, 
for each cluster \( C_i \), the \( m \) samples closest to the cluster centroid \( \mu_i \) are selected as representative samples \( S_i \) (Lines 12--13). 
These samples are then annotated using an LLM (Line 14), and the final cluster label \( Y_{C_i} \) is determined via majority voting over their labels \( Y_{S_i} \), which is then propagated to the rest of the cluster (Line 15).

After determining the cluster label, the label is propagated to all other log samples within the cluster \( E_j \in C_i \), ensuring that structurally similar log messages receive consistent labels (Line 16). This propagation process results in the construction of an augmented annotated dataset where each log embedding \( E_j \) is assigned the corresponding cluster label \( Y_{C_i} \) (Lines 17--19). The resultant dataset is formally defined as:
\begin{equation}
    \mathcal{D}_{\text{labeled}} = \{(E_j, Y_{C_i}) \mid E_j \in C_i, \forall C_i \in C\}.
\end{equation}
Particularly, the combination of majority voting and clustering-based label propagation enhances label robustness by mitigating the impact of outliers and ensures consistent labeling of structurally similar logs without requiring any human intervention.

\subsubsection{Two-Step Few-Shot Refinement for LLM-Based Labeling}
As shown in Table~\ref{llm}, using randomly selected FS examples leads to limited labeling performance, which is insufficient for directly annotating log data.
To enhance the reliability and robustness of LLM-based log annotation, we propose a two-step refinement framework for FS prompt construction. 
In the first step, we perform an initial evaluation of the LLM's annotation performance using a small validation set, where the prompt includes randomly selected FS log samples. We then identify the subset of validation samples that are incorrectly predicted. Those misclassified samples are treated as the hard-annotated cases and are incorporated into the prompt by replacing some of the initial randomly selected FS examples. 

In the second step, we analyze the error distribution across different semantic categories or log types. Specifically, we identify the clusters that are frequently misclassified and selectively add a small number of representative log samples from these categories to the FS prompt set. This targeted augmentation helps the LLM focus on ambiguous or challenging log types that are underrepresented in the initial prompt.
We note that the validation set in this pre-analysis stage is a small but manually annotated subset prepared prior to ALPHA pipeline. This one-time effort is only for refining FS prompts at the beginning, after which the pipeline operates fully automatically without human intervention.
Experimental results in Sec.~\ref{performance} demonstrate that such refinement strategy significantly enhances the labeling accuracy.

\subsection{Log Anomaly Detection and Analysis}

\subsubsection{Training the Detector with Augmented Data}

Once log messages are labeled through our clustering-based propagation, an anomaly detector is trained to identify abnormal patterns within the network system. Classical ML models, such as SVM and decision trees, have demonstrated strong accuracy and efficiency in log-based anomaly detection tasks~\cite{yu2024deep}. Consequently, more advanced techniques like deep learning are not always necessary, especially when labeled data is sufficient and deployment costs are a concern. In our work, we adopt classical ML models for their lightweight nature; however, our framework remains model-agnostic and can be readily extended to incorporate more sophisticated deep learning architectures for log analysis if needed.

Due to the high computational cost associated with Transformer-based log embeddings, which are often unsuitable for deployment on resource-constrained devices, we propose using an augmented feature set that balances performance with efficiency, combining both \textbf{semantic features} and \textbf{text-based features}. 
Specifically, the semantic features are extracted to capture high-level attributes of log messages, including the presence of error-related keywords, message length, numerical patterns, and references to system paths, IP addresses, and ports. These structured attributes offer interpretable insights into potential anomalies by identifying patterns commonly associated with system failures or security incidents. In contrast, the text-based features provide a more flexible representation of the log messages by employing text vectorization techniques such as {TF-IDF}~\cite{salton1988term}. This method encodes the logs into numerical vectors based on word occurrence statistics, allowing the detection model to generalize across diverse log formats without relying on predefined rules.

To ensure robustness and minimize false positives, a predefined threshold is set for model performance validation in the proposed pipeline. If the threshold is not met, the detection model undergoes an iterative refinement, where additional log data is collected and incorporated into the training set. The pipeline is subsequently updated using our active learning strategy to select the most informative samples, enabling the dataset to grow dynamically with high-quality labeled logs. This iterative process continues until the detection performance meets the desired threshold, ensuring an adaptive and continuously improving anomaly detection system.
By integrating the augmented feature set, iterative model refinement, and efficient ML classifiers, ALPHA achieves robust and scalable log anomaly detection, enabling effective identification of abnormal system behaviors with minimal manual intervention.

\begin{figure}[t]
    \centering
    \begin{tcolorbox}[
        colframe=black!75!white, 
        colback=gray!5, 
        sharp corners=southwest, 
        boxrule=0.8pt, 
        width=0.95\linewidth,
        fontupper=\scriptsize
    ]
        \textbf{System Prompt:} \\
        \textit{You are a cybersecurity analyst specializing in anomaly detection. 
        Your task is to analyze a given log entry that has been flagged as anomalous. 
        Identify the \textbf{top 2--3 most probable root causes} and provide \textbf{concise} recommendations for further investigation. 
        Keep your response \textbf{brief and to the point}, avoiding unnecessary details. 
        Limit the response to \textbf{a maximum of 100 words}.}

        \vspace{0.2cm}
        \textbf{User Prompt:} \\
        \textit{Analyze the following anomalous log entry:} \texttt{[log\_entry]}. \textit{Provide:}
        \begin{itemize}
            \item \textit{The possible causes of the anomaly.}
            \item \textit{Recommendations for further action.}
        \end{itemize}
    \end{tcolorbox}
    \caption{System and User Prompts used for LLM-based Anomaly Explanation.}
    \label{llm_anomaly_prompt}
    \vspace{-6mm}
\end{figure}

\subsubsection{LLM-Enabled Anomaly Analysis}
\label{sec:prompting}
Existing anomaly detection models are effective at identifying anomalies in log data but lack interpretability. These models typically output binary or probabilistic labels indicating whether a log entry is anomalous and do not provide insights into the underlying causes of anomalies. As a result, domain experts are required to manually investigate anomalous logs, analyze patterns, and identify root causes, which is a time-consuming and labor-intensive process. 

To address this limitation, we incorporate a zero-shot LLM-based Root Cause Analysis (RCA) module into ALPHA that automatically generates concise, structured, and interpretable diagnostic reports for detected anomalies. As illustrated in Fig.~\ref{llm_anomaly_prompt}, we formulate a carefully crafted system and user prompt that instructs the LLM to behave as a cybersecurity analyst. The prompt explicitly requests the model to identify the top 2–3 most probable root causes and provide actionable recommendations in no more than 100 words.

This zero-shot prompting approach requires no labeled examples and relies solely on task-specific instructions to elicit accurate and consistent responses. The LLM effectively transforms low-level system logs into human-readable explanations, significantly reducing the burden on domain experts. Additionally, the structured format enables downstream integration with automated incident management systems. By leveraging log-specific language patterns and well-designed prompts, the LLM significantly reduce the manual effort required for RCA purposes. Furthermore, once the root causes are identified, the LLM can be extended to interact with existing remediation tools to assist in resolving the detected issues.

Overall, by assigning LLMs well-defined roles across log data embedding, clustering-based active sampling, and the RCA process, ALPHA achieves truly human-free anomaly detection throughout the entire system lifecycle.

\section{Experiments and Evaluations}

\subsection{Experimental Settings}
\subsubsection{Foundation models} Several well-known pretrained text embedders have been developed such as BERT \cite{devlin2019bert} and RoBERTa \cite{liu2019roberta}. These models are trained on large-scale corpora to learn rich contextual representations, making them effective for log data embedding.
In our experiments, the text-embedding-ada-002 \cite{neelakantan2022text} model provided by OpenAI is used to generate semantic embeddings of log messages for clustering and downstream analysis. In addition, both the LLM-based labeling and RCA process in ALPHA are powered by OpenAI's GPT-4o model \cite{hurst2024gpt}.

\subsubsection{Datasets}
We utilize the Thunderbird dataset \cite{oliner2007supercomputers}, a system log dataset from a large-scale data center network environment, to evaluate the effectiveness of ALPHA for automated log anomaly detection and interpretation.
The dataset consists of both benign and anomalous log entries. 
In the log anomaly detection task, the distribution of anomalies significantly impacts detection performance. A lower anomaly rate may lead to high false negative rates, as models tend to favor the majority class, whereas a higher anomaly rate can make the task artificially easier. To assess the robustness of our approach, we design two test sets with different anomaly rates, as shown in Table~\ref{dataset}. \textbf{Imbalanced Test Set (I)} with an anomaly rate of approximately 20\%, while \textbf{Balanced Test Set (B)} has a higher anomaly rate of 40\%. Then, we can examine how well our proposed ALPHA generalizes under varying anomaly distributions.

\subsubsection{Evaluation metrics}
We adopt two standard metrics commonly used in network anomaly detection: Accuracy and F1-score. Accuracy reflects the model’s overall effectiveness in classifying both normal and anomalous events, while the F1-Score captures the trade-off between false positives and false negatives. {We also quantify the human-in-the-loop cost by measuring the number of manual annotations required under different label sources.}

\begin{table}[t]
\centering
\caption{Network Log Datasets.}
\begin{tabular}{c|cc}
\toprule
\textbf{Dataset} & \textbf{Benign Logs} & \textbf{Anomalous Logs} \\
\midrule
Training Set & 4,000 & 6,000 \\
Imbalanced Test Set (I)& 3,910  & 1,090 \\
Balanced Test Set (B) & 3,000 & 2,000 \\
\bottomrule
\end{tabular}
\label{dataset}
\end{table}

\subsection{Performance of ALPHA}
\label{performance}
\subsubsection{\textbf{Cluster Optimization}}
\begin{figure}[t] 
    \centering
    \begin{subfigure}{0.47\linewidth}
        \centering
        \includegraphics[width=\linewidth]{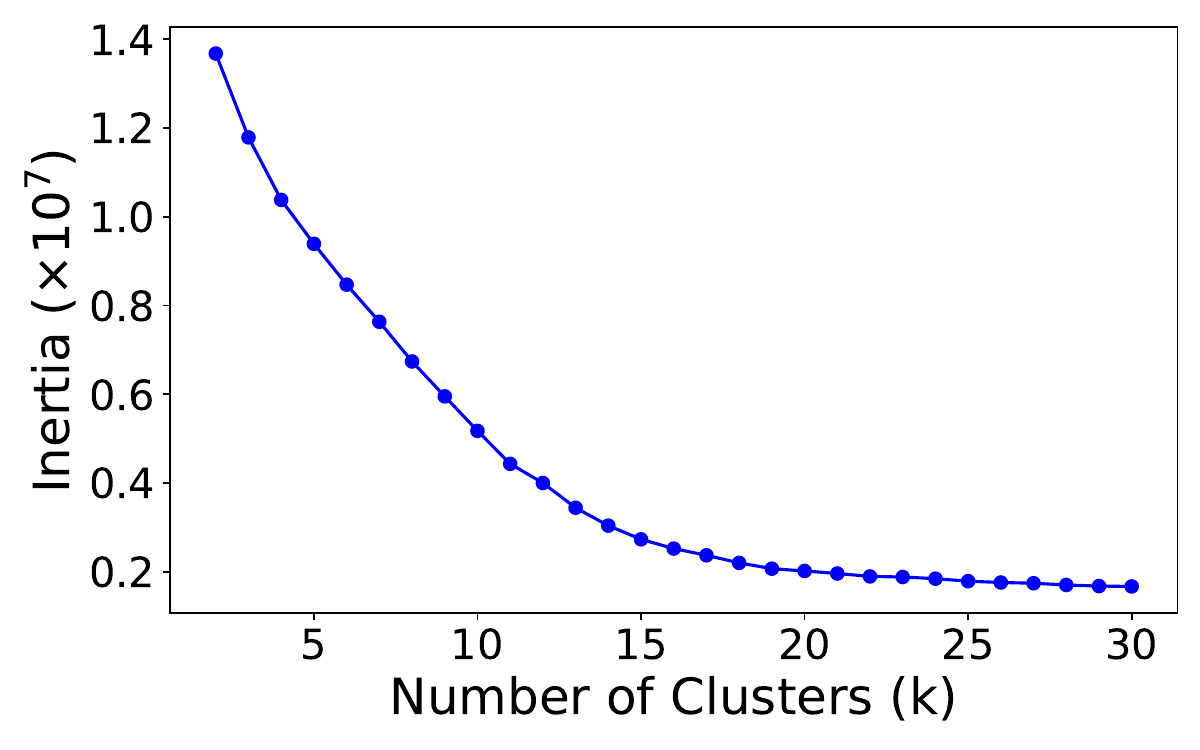}
        \caption{Elbow Method.} 
	\label{Elbow}
    \end{subfigure}
    \hfill
    \begin{subfigure}{0.47\linewidth}  
        \centering
        \includegraphics[width=\linewidth]{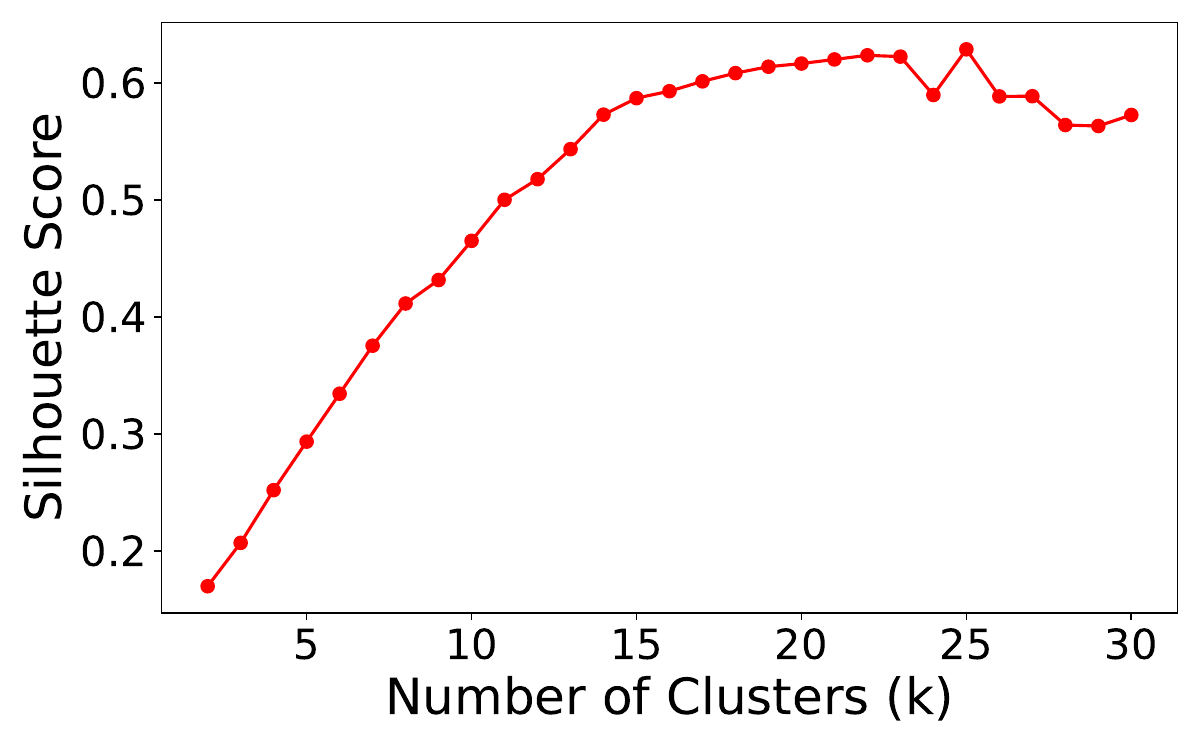}
        \caption{Silhouette Analysis.} 
	\label{Silhouette}
    \end{subfigure}
    
    \caption{Cluster Optimization Results.} 
	\label{cluster}
    \vspace{-5mm}
\end{figure}

To ensure effective label propagation and accurate representation of log structure, we first optimize the number of clusters used in our k-means clustering module, i.e the joint Elbow method and Silhouette analysis, as illustrated in Fig.~\ref{cluster}.
While WCSS (Inertia) naturally decreases with more clusters, the rate of decline significantly slows down after an elbow point. This suggests a balance between clustering compactness and model complexity. As shown in Fig.~\ref{Elbow}, the elbow appears around $k=15$, beyond which further increases in cluster count yield diminishing returns.
The silhouette analysis further supports this decision by measuring how similar a point is to its own cluster compared to others. As shown in Fig.~\ref{Silhouette}, the average silhouette score peaks near $k=15-20$, with $k=15$ achieving a strong trade-off between intra-cluster cohesion and inter-cluster separation. Based on this joint evaluation, we choose $k=15$ as the optimal number of clusters for the subsequent label propagation process.

\subsubsection{\textbf{Performance of Labeling}} 
\begin{figure}[t]  
    \centering
    \begin{subfigure}{0.49\linewidth}
        \centering
        \includegraphics[width=\linewidth]{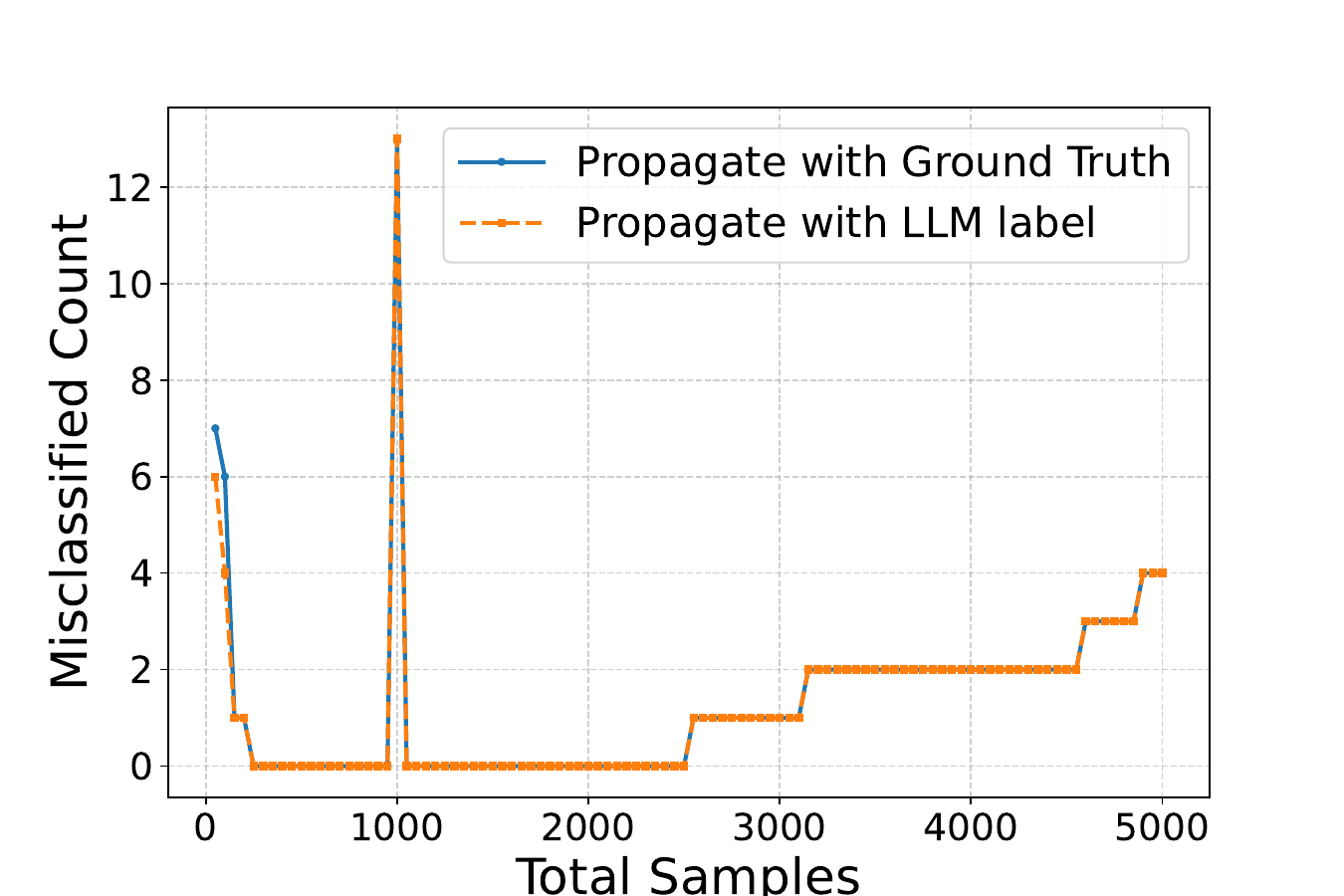}
        \caption{Misclassified Count with Different Propagation Methods.} 
	\label{mis count}
    \end{subfigure}
    \hfill
    \begin{subfigure}{0.49\linewidth}  
        \centering
        \includegraphics[width=\linewidth]{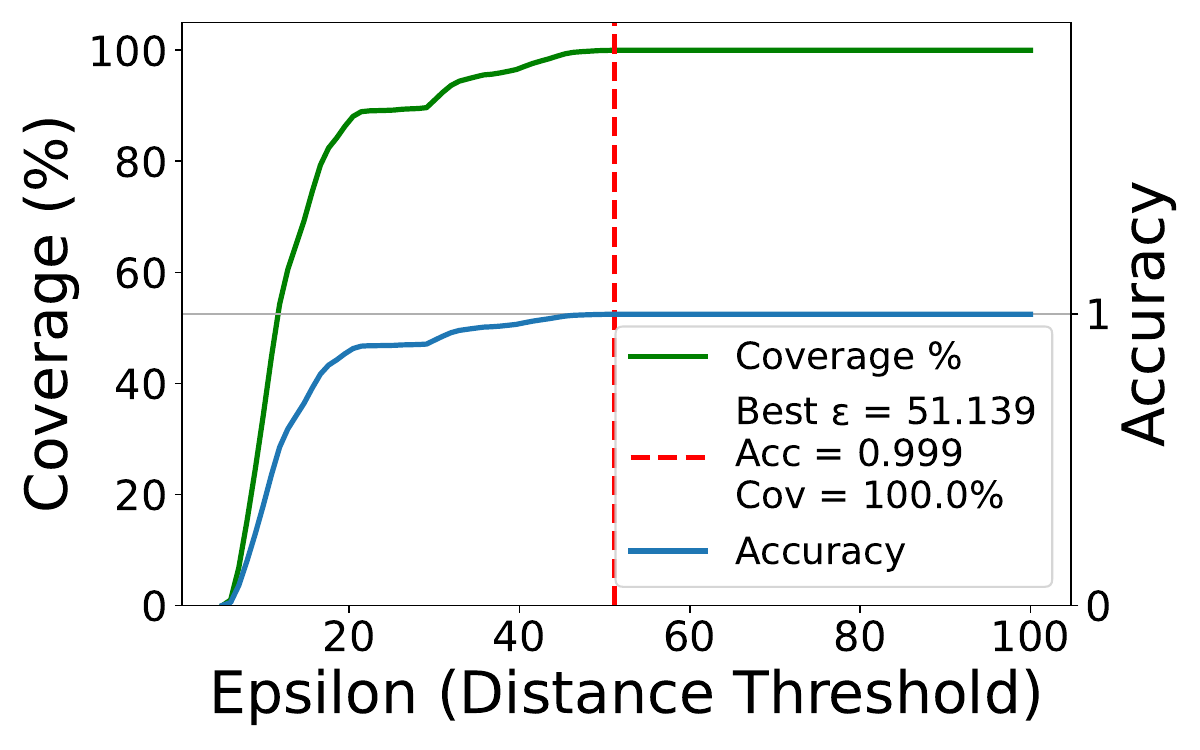}
        \caption{Accuracy and Coverage vs. Flipping Distance.} 
	\label{flip}
    \end{subfigure}
    
    \caption{Performance of Cluster-based Labeling.} 
	\label{cluster performance}
\end{figure}

First, to validate the accuracy of our cluster-based labeling strategy, we compare two label propagation methods: one using the ground-truth labels of the data points nearest to the cluster center, and another using the majority vote results from the LLM annotations on the top-5 closest points.
As shown in Fig. \ref{mis count}, both approaches exhibit highly similar performance across varying sample sizes, with nearly identical misclassification counts. Remarkably, when the total number of samples is small, e.g., 50 or 100, the LLM-based propagation occasionally outperforms the ground-truth method. This is because the majority voting mechanism from LLM-labeled samples mitigates noise that may exist when some clusters may only contain a few samples.
Aside from the outlier at around 1,000 samples, misclassification counts for all other sample sizes remain extremely low and increase gradually. Given that the maximum number of misclassified samples remains under 5 across 5,000 total log samples, the result demonstrates the high quality of our cluster assignment and label propagation method.
Combined with our two-step few-shot selection strategy and LLM-based majority voting, this pipeline proves to be robust and reliable, providing an effective foundation for human-free log labeling at scale.

Second, to evaluate the reliability of cluster-based annotation, we introduce a distance threshold \( \varepsilon \) to examine whether flipping the predicted label for log samples far from the cluster center can enhance annotation accuracy. For each sample, we compute its distance to the assigned cluster center and reverse its predicted label if the distance exceeds \( \varepsilon \).
We perform a sweep over a range of \( \varepsilon  = 0-100\) and report the resulting annotation accuracy and coverage using a small labeled validation set. As shown in Fig. \ref{flip}, the coverage represented by the green line quickly rises and reaches $100\%$ near \( \varepsilon =51\). Meanwhile, the cluster accuracy shown by the blue line improves with increasing \( \varepsilon \), reaching a peak of $99.9\%$ when full coverage is achieved. At this point, no labels are flipped, meaning that the original cluster labels are directly used without modification. This suggests that the initial cluster labels are already highly reliable, and additional label flipping is unnecessary. Based on this result, we choose not to apply additional label flipping in subsequent experiments and directly use the initial cluster labels as propagated annotations.

\begin{figure}[t] 
    \centering
    \begin{subfigure}{0.325\linewidth}  
        \centering
        \includegraphics[width=\linewidth]{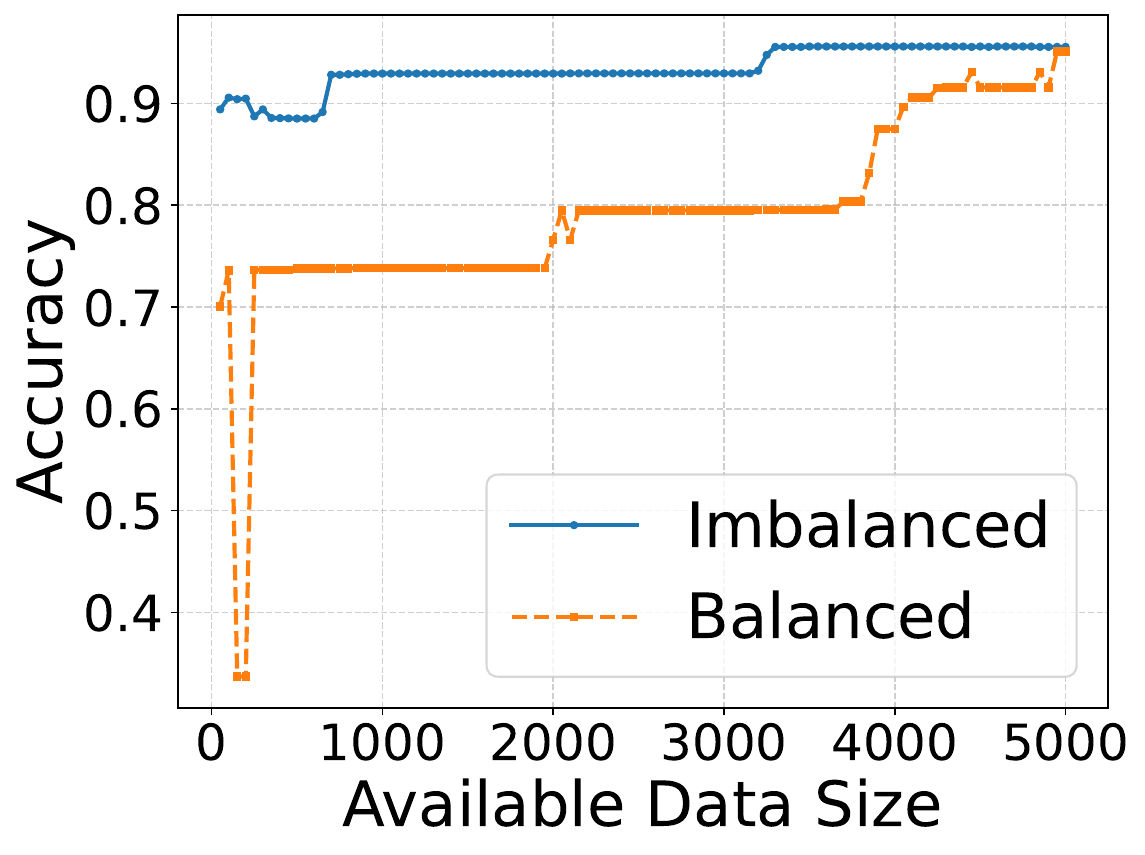}
        \caption{Supervised learning with labels} 
	\label{true}
    \end{subfigure}
    \hfill
    \begin{subfigure}{0.325\linewidth}
        \centering
        \includegraphics[width=\linewidth]{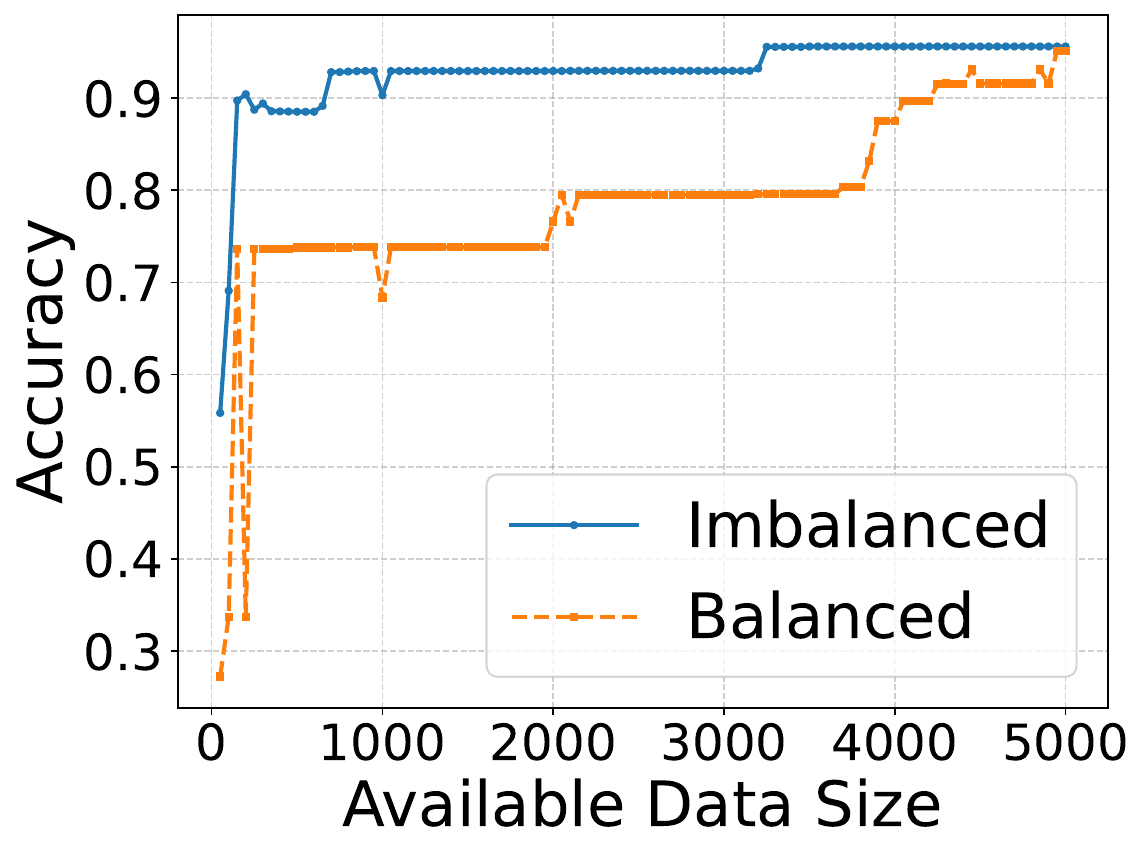}
        \caption{Active learning with annotated labels} 
	\label{true pro}
    \end{subfigure}
    \hfill
    \begin{subfigure}{0.325\linewidth}
        \centering
        \includegraphics[width=\linewidth]{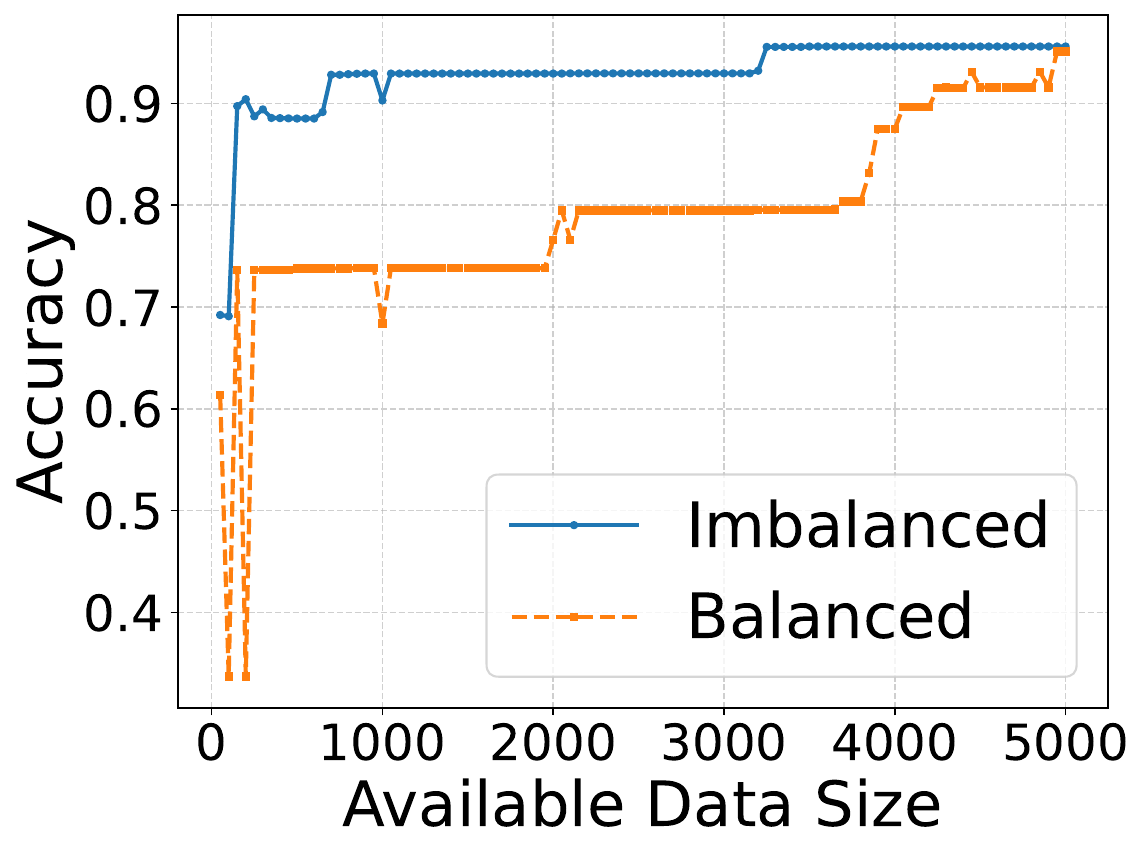}
        \caption{ALPHA (ours)} \vspace{3.5mm}
	\label{llm pro}
    \end{subfigure}
    \caption{Anomaly detection performance with increasing available data during active learning across different label sources.}
    \vspace{-3mm}
    \label{acc}
\end{figure}

\subsubsection{\textbf{Log Anomaly Detection Performance}}
We evaluate the downstream anomaly detection performance using a SVM trained on three different label sources with increasing available data collected by the system. Fig. \ref{acc}(a) shows the detection accuracy of the model trained with human-annotated (ground-truth) log samples. Fig. \ref{acc}(b) presents the result of active learning on labels propagated using cluster centers with human-annotated logs, while Fig. \ref{acc}(c) corresponds to our human-free ALPHA pipeline.
Overall, we observe that all three approaches exhibit consistent accuracy trends across varying available data sizes. Specifically, the models trained on propagated labels (Fig. \ref{true pro} and Fig. \ref{llm pro}) show slightly more fluctuations than the fully supervised baseline (Fig. \ref{true}), especially when the available data size is small or around 1,000 samples. This aligns with our previous findings {(in Fig.~\ref{mis count})} that misclassification counts spike at the 1,000-sample mark due to noisy cluster assignments. 
Nevertheless, the fluctuations in the active learning scenarios remain relatively minor, and accuracy steadily improves as available size increases. This shows that small amounts of mislabeling in propagated labels do not impact downstream classifier performance. 

\begin{table}[t]
\centering
\caption{F1-score and Human Annotation Cost Across Different Strategies.}
\begin{tabular}{c|c|cc|c}
\toprule
\textbf{Label source} & \textbf{Test Set} & \textbf{Model} & \textbf{F1 (\%)} & \textbf{\#Annotation Cost} \\
\midrule
\multirow{4}{*}{\textbf{\shortstack{Supervised \\ learning  with \\ human labels}}} 
& \multirow{2}{*}{I} & SVM & 95.70 & \multirow{4}{*}{\textbf{5,000}} \\
& & LR & 99.19 & \\
\cmidrule(l){2-4}
& \multirow{2}{*}{B} & SVM & 95.15 & \\
& & LR & 99.44 & \\
\midrule
\multirow{4}{*}{\textbf{\shortstack{Active learning \\ with \\ human labels}}} 
& \multirow{2}{*}{I} & SVM & 95.74 & \multirow{4}{*}{\shortstack{\textbf{\boldmath{$k \times n$}}}} \\
& & LR & 99.19 & \\
\cmidrule(l){2-4}
& \multirow{2}{*}{B} & SVM & 95.16 & \\
& & LR & 99.34 & \\
\midrule
\multirow{4}{*}{\textbf{\shortstack{ALPHA (ours)}}} 
& \multirow{2}{*}{I} & SVM & 95.74 & \multirow{4}{*}{\textbf{0}} \\
& & LR & 99.19 & \\
\cmidrule(l){2-4}
& \multirow{2}{*}{B} & SVM & 95.16 & \\
& & LR & 99.34 & \\
\bottomrule
\end{tabular}
\label{table_f1}
\end{table}

Regarding the detection accuracy in Fig.~\ref{acc}, when the available data reaches 5,000 samples, we observe that the accuracy on both test sets consistently exceeds 95\%. To further investigate the effectiveness of different labeling strategies under this data scale, we compare the anomaly detection performance of classifiers trained with 5,000 available data samples but labeled through varying levels of human effort, as summarized in Table~\ref{table_f1}.
Beyond the SVM model, we also adopt a Logistic Regression (LR) model to showcase the \textit{generalizability} of our pipeline, ensuring its effectiveness regardless of the specific ML-based detector employed.
As shown in Table~\ref{table_f1}, both models achieve consistently high F1-scores across all label sources and test sets. 
Notably, the fully supervised requires default 5,000 human annotations, while the active learning variant reduces this effort to $k \times n$ annotations, where $k$ is the number of clusters and $n$ is the number of samples closest to the cluster center used for majority voting (e.g. $15 \times 1$ in our setting). Most importantly, our ALPHA pipeline achieves nearly identical detection performance, over 95\% F1-score for SVM and over 99\% for LR, while requiring zero human annotation cost.
\textit{These results demonstrate that the proposed ALPHA pipeline, while fully human-free, achieves scalable and reliable anomaly detection in log data.}

\subsubsection{\textbf{LLM-based Log Anomaly Analysis}}
In addition to reaching a predefined performance threshold, our pipeline supports in-depth log interpretation through LLM-based analysis. As described in Sec.~\ref{sec:prompting}, a structured prompt is designed to guide the LLM to analyze anomalous logs by identifying root causes and suggesting remediation actions.
Fig.~\ref{log_analysis} shows a representative example of this capability. The input log, which contains connection refusal errors from the \texttt{pbs\_mom} process, is automatically interpreted by the LLM. The analysis identifies two possible causes, service misconfiguration or network policy restrictions, and provides actionable recommendations, such as verifying the service status, reviewing network policies, and checking surrounding logs for related errors.
This RCA function shows the LLM's potential not only to detect but also to \textit{contextualize} anomalies, bridging the gap between detection and explainability. More importantly, the entire process is conducted without domain expert intervention, demonstrating the effectiveness of our LLM's dual role in enabling both anomaly labeling and network analysis in a truly human-free manner.

\begin{figure}[t]
    \centering
    \begin{tcolorbox}[
        colframe=black!75!white, 
        colback=gray!5, 
        sharp corners=southwest, 
        boxrule=0.8pt, 
        width=0.95\linewidth,
        fontupper=\scriptsize
    ]
        \textbf{Log Entry:} \\
        \texttt{...,pbs\_mom, Connection refused (111) in open\_demux, 
        open\_demux: cannot connect to 10...126:45841,..., 
        Connection refused (<*>) in open\_demux, open\_demux: cannot connect to <*>, 
        ['111', '10...126:45841']} 

        \vspace{0.1cm}
        \textbf{LLM Analysis:}

        \textbf{1. Possible Cause of the Anomaly}
        \begin{itemize}
            \item \textbf{Service Down or Misconfigured}: The service on IP \texttt{10...126} at port \texttt{45841} may be down or incorrectly configured, causing connection refusals.
            \item \textbf{Network Issues}: Firewall or network policies might be blocking connections to the specified IP and port.
        \end{itemize}

        \textbf{2. Recommendations}
        \begin{itemize}
            \item \textbf{Verify Service Status}: Check if the service on \texttt{10...126:45841} is running and properly configured.
            \item \textbf{Check Network Policies}: Review firewall rules and security settings for possible blocks or misconfigurations.
            \item \textbf{Analyze Logs}: Investigate logs around the timestamp for any related network or service errors.
        \end{itemize}

    \end{tcolorbox}
    \vspace{-3mm}
    \caption{Analysis of Anomalous Log Entry.}
    \label{log_analysis}
    \vspace{-5mm}
\end{figure}

\section{Conclusion}

In this paper, we propose ALPHA, a novel LLM-enabled active learning framework designed to achieve fully human-free log-based anomaly detection and analysis. ALPHA integrates semantic embedding, clustering-based sampling, LLM-assisted annotation, and iterative model refinement into a unified pipeline. 
Experimental results demonstrate that models trained on LLM-propagated labels achieve comparable accuracy to those trained on human-annotated data across different test sets and classifiers. This is attributed to our two-step few-shot refinement strategy, which enhances LLM annotation quality and ensures that downstream classifiers remain robust to minor labeling noise. Furthermore, ALPHA extends beyond detection by incorporating LLM-based explanations for root cause analysis, delivering both automation and interpretability. This fully human-free pipeline advances scalable and cost-effective log analysis for real-world network systems.

\section*{Acknowledgment}

This research was supported by the National Science Foundation through Award CNS--2440756, CNS--2312138 and SaTC-2350075.
\vspace{-1mm}
\bibliographystyle{IEEEtran}
\bibliography{reference}

\end{document}